\documentclass[aps,showpacs,amsmath,amssymb,twocolumn,prb,floatfix,superscriptaddress,a4]{revtex4}
\usepackage{graphicx}
\usepackage{epsfig}
\usepackage{amsmath}
\usepackage{amsfonts}
\usepackage{amssymb}
\usepackage{color}

\newcommand{\abs}[1]{\left\lvert#1\right\rvert}
\newcommand{\ket} [1] {\vert #1\rangle}
\newcommand{\bra} [1] {\langle #1\vert}

\newcommand{\expect}[1]{\langle #1\rangle}
\begin{document}
\title{Confinement sensitivity in quantum dot singlet-triplet relaxation}

\author{C. J. Wessl{\'e}n}
\affiliation{Department of Physics, Stockholm University, AlbaNova, S-106
91 Stockholm, Sweden}

\author{E. Lindroth}
\affiliation{Department of Physics, Stockholm University, AlbaNova, S-106
91 Stockholm, Sweden}

\date{\today}

\begin{abstract}
Spin-orbit mediated phonon relaxation in a two-dimensional quantum dot is
investigated using different confining potentials. Elliptical harmonic
oscillator and cylindrical well results are compared to each other in the case
of a two-electron GaAs quantum dot subjected to a tilted magnetic field. The
lowest energy set of two-body singlet and triplet states are calculated
including spin-orbit and magnetic effects. These are used to calculate the
phonon induced transition rate from the excited triplet to the ground state
singlet for magnetic fields up to where the states cross. The roll of the cubic
Dresselhaus effect and the positioning of "spin hot-spots" are discussed and
relaxation rates for a few different systems are exhibited.
\end{abstract}

\maketitle
\section{Introduction\label{sec:intro}}
Quantum dots (QDs) created through semiconductor heterostructures are
promising components in the fields of spintronics, quantum information and
quantum computing~\cite{RevModPhys.76.323,PhysRevA.57.120}. For time sensitive
applications, as when operations are to be performed on excited electron states
within their coherence time, a long relaxation time is a decisive  property.
Long lifetimes are expected if the decay requires a spin flip and experimental
studies have here measured excited state relaxation on the  microsecond to
millisecond timescale~\cite{Kroutvar2004,Fujisawa2002,
Elzerman2004,PhysRevLett.94.196802}, with the longest life times for dots
subjected to a magnetic field. It is important to understand, and to
quantitatively model, the dominating electronic and spin-dependent relaxation
processes.

The role of the magnetic field is to suppress
direct spin exchange with bulk materiel nuclei~\cite{Fujisawa2002}.
The still occurring relaxation is  in this situation believed to be dominated
by phonon exchange, but this mechanism opens the route to spin flips only in
combination with spin-orbit coupling. This has  been an important area of study,
both for single electron \cite{NewJPhys.14.013035,PhysRevB.87.235202,
PhysRevB.74.035313} and multi electron states \cite{PhysRevB.77.115438,
PhysRevLett.94.196802, PhysRevLett.98.126601,PhysRevLett.113.256802}, where the
role of the Rashba\cite{rashba} as well as the Dresselhaus\cite{dresselhaus}
mechanisms for spin-orbit interaction in semiconductors have been investigated.

For two electron GaAs dots in particular,
relaxation times on the millisecond timescale have been demonstrated
experimentally ~\cite{PhysRevLett.94.196802, PhysRevLett.98.126601}, and
the singlet-triplet energy-splitting, as well as  relaxation rates, have been
investigated in some detail for magnetic fields of up to a few Tesla.
Previous theoretical
efforts to model these
studies have shown that the system seems to exhibit a smaller than expected
Dresselhaus coefficient~\cite{PhysRevB.85.035326,PhysRevB.76.085305} and that a
cylindrical hard-wall potential reproduces the singlet-triplet energy-splitting
more accurately than a harmonic oscillator potential~\cite{Frostad2013}. In the
present work, an elliptical cylinder potential is used to further study the
realistic parameter range determining the energy splitting and relaxation
rates of said system, motivated by the better energy-splitting agreement
previously shown. Harmonic oscillator results are also produced using the same
methods and the differences in results due to potential shape are investigated.

This paper starts with a brief model- and implementation section describing what
physical effects are included and how they are implemented. The electron
wave-functions are represented in a B-spline basis~\cite{deboor}, allowing for
an arbitrary potential shape. Magnetic effects, in the form of a tilted magnetic
, as well as spin-orbit (SO) effects are included in the one-body Hamiltonian.
All the one-particle contributions to the Hamiltonian are used to create a basis
which in turn is used to include the Coulomb interaction through the full
configuration interaction (FCI) method. The two-electron states corresponding to
the singlet ground state and first excited
triplet states are extracted and used to calculate the phonon induced relaxation
rate from Fermi's golden rule. Dot width, ellipticity and
relative directions of the tilted magnetic field are varied to match the energy
splitting profile from the experimental results.

Following the model and implementation section, the results show the calculated
singlet-triplet energy splittings for different dot potentials and compared them
to the experimental results by Meunier et al.\cite{PhysRevLett.94.196802}.
This is followed by relaxation rates calculated for the same systems.

\section{Model and Implementation\label{sec:method}}
\subsection{Model}
The two electron Hamiltonian describing the quantum dot can be expressed as
\begin{equation}
\label{2eHam}
H = \sum_{i=1,2} h(\mathbf{r}_i) +
\frac{e^2}{4\pi\epsilon_r\epsilon_0 \abs{\mathbf{r}_1-\mathbf{r}_2}},
\end{equation}
where $\epsilon_r$ is the relative permittivity of the semiconductor material
and  $h(\mathbf{r}_i)$ is the one-particle Hamiltonian:
\begin{equation}
\label{1eHam}
h(\mathbf{r}_i) =\frac{1}{2m^*}\mathbf{\hat{\Pi}}^2+
g^*\mu_b \mathbf{B}\cdot \mathbf{\hat{S}}+V+h_{SO}.
\end{equation}
Here $\mathbf{\hat{\Pi}}$ is the momentum operator, $\mathbf{\hat{S}}$ is the
spin operator vector, $V$ is the effective confinement potential and
$h_{SO}$ is the spin-orbit interaction.
Bulk material properties are used for the electron effective mass $m^*$ and
gyromagnetic ration $g^*$.
The magnetic field $\mathbf{B}$ is tilted from the z-axis by an angle $\theta$
and azimuthally from the x-axis by $\phi$, so that:
\begin{equation}
\label{B}
\mathbf{B}=B_0(\cos\phi \sin\theta,\sin\phi \sin\theta,\cos\theta),
\end{equation}
and the kinetic energy operator can be expanded to:
\begin{equation}
\label{kin}
\frac{1}{2m^*}\mathbf{\hat{\Pi}}^2=-\frac{\hbar^2}{2m^*}\nabla^2+
\frac{e^2}{2m^*}A^2+\frac{e}{2m^*}\mathbf B \cdot \mathbf{\hat{L}}.
\end{equation}

Here we choose to study a QD restricted in the z-dimension, often orders of
magnitude thinner than wide. We can divide the wave function as:
\begin{equation}
\label{wfz}
\Psi(\mathbf r)= \psi(x,y)\delta(z),
\end{equation}
where the z-component is assumed to be in the shape of a Dirac delta function.
This limits the vector potential to:
\begin{equation}
\label{A}
\mathbf A= \frac{B_0}{2}(-y \cos\theta,x \cos\theta,
y \cos\phi \sin\theta-x \sin\phi\sin\theta),
\end{equation}
and with these restrictions $A^2$ takes the form of an
anisotropic harmonic oscillator potential in the xy-plane.
The potential has an elliptical cross section and the minor axis lies
along  projection of the magnetic field in the xy-plane (i.e. in the
the $\phi$
direction),
\begin{equation}
\label{A2}
A^2=\frac{B_0}{4} \left[(x^2+y^2)\cos^2\theta +
(y\sin\phi-x\cos\phi)^2\sin^2\theta\right].
\end{equation}

The angular momentum operator will also for this system be limited to
$\mathbf{\hat{L}}=\hat{L_z}$ due to the restraints in the z-direction, leaving
the angular momentum Zeeman term:
\begin{equation}
\label{BL}
\frac{e}{2m^*}\mathbf{B}\cdot\mathbf{\hat{L}}=
\frac{e}{2m^*}\cos\theta B_0 \hat{L_z}.
\end{equation}

With $\mathbf{B}$ placed along the z-axis, the electron spin states are
eigenstates to the $\hat{S_z}$ operator, leading to the
Zeeman spin term shifting the states $\ket{\pm1/2}$ by the energy:
\begin{equation}
\label{zeeman_z}
\mp g^*\mu_b B_0\hbar/2.
\end{equation}
With an inclined magnetic field the ${\hat{S_z}}$ eigenstates will no longer be
eigenstates to the Zeeman operator, instead the Zeeman spin term will couple the
$\ket{\pm1/2}$ states by:
\begin{equation}
\label{zeeman}
\begin{gathered}
g^*\mu_b \mathbf{B}\cdot \mathbf{\hat{S}}=\\
g^*\mu_b B_0\left(\cos\phi\sin\theta\hat{S_x}+
\sin \phi\sin\theta\hat{S_y}+\cos\theta\hat{S_z} \right).
\end{gathered}
\end{equation}
Note that this alone will not mix spins, but rather lead to spin up and down
states (as defined by the direction of the magnetic field) that are linear
combinations of the $\hat{S_z}$ eigenstates according to:
\begin{equation}
\ket{\uparrow}=\cos(\theta/2)\ket{+1/2}-\sin(\theta/2)\ket{-1/2},
\end{equation}
and
\begin{equation}
\ket{\downarrow}=\sin(\theta/2)\ket{+1/2}+\cos(\theta/2)\ket{-1/2},
\end{equation}
with the same energy splitting as without any field inclination.

Spin mixing can occur through the spin-orbit Hamiltonian,
$h_{SO}=H_R+H_{D1}+H_{D3}$, including  the Rashba:
\begin{equation}
\label{R}
h_R=\frac{\alpha}{\hbar}(\hat{\pi}_x\hat{\sigma}_y-\hat{\pi}_y\hat{\sigma}_x),
\end{equation}
the linear Dresselhaus:
\begin{equation}
\label{D}
h_{D1}=\frac{\gamma}{\hbar^3}\expect{\hat{\pi}_z^2}(\hat{\pi}_y\hat{\sigma}_y-
\hat{\pi}_x\hat{\sigma}_x),
\end{equation}
and the cubic Dresselhaus:
\begin{equation}
\label{D3}
h_{D3}=\frac{\gamma}{2\hbar^3}[(\hat{\pi}_x \hat{\pi}_y^2\hat{\sigma}_x-
\hat{\pi}_y \hat{\pi}_x^2\hat{\sigma}_y)+
(\hat{\pi}_x \hat{\pi}_y^2\hat{\sigma}_x-
\hat{\pi}_y \hat{\pi}_x^2\hat{\sigma}_y)^\dagger].
\end{equation}
where $\alpha$ and $\gamma$ are the Rashba and Dresselhaus coefficients,
$\hat{\pi}_{x,y,z}$ are the momentum operators and $\hat{\sigma}_{x,y}$ are the
Pauli spin matrices. The Rashba interaction conserves $m_l+m_s$ and linear
Dresselhaus interaction conserves $m_l-m_s$. The cubic Dresselhaus operator
couple states $\ket{m_l,m_s}$ according to the selection
rules\cite{PhysRevB.72.155410}:
\begin{equation}
\label{selectionrules}
\begin{split}
\ket{m_l,+1/2}\rightarrow \ket{m_l-1,-1/2},\ket{m_l+3,-1/2}\\
\ket{m_l,-1/2}\rightarrow \ket{m_l+1,+1/2},\ket{m_l-3,+1/2}.
\end{split}
\end{equation}

Previous studies on similar systems have shown the Dresselhaus coupling to be
the dominating spin-orbit interaction in this regime\cite{PhysRevB.64.125316},
with the cubic Dresselhaus term being of some significance
\cite{PhysRevLett.98.226802}. The ratio between the Dresselhaus and
Rashba coefficients may however still be of interest when studying an
anisotropic QD under a tilted and rotated magnetic
field\cite{PhysRevLett.113.256802}, small Rashba coefficients have been tested
and found not to change the results significantly and will not be investigated
further in this paper.

In this work we focus on two effective confinement potentials; the two
dimensional harmonic oscillator
\begin{equation}
\label{HO}
V_{HO}(\mathbf r_i)=\frac{m^*\omega^2}{2}[(\delta x_i)^2+(\frac{y_i}{\delta})^2],
\end{equation}
where $\omega$ is the harmonic oscillator frequency and $\delta$ is the dot
ellipticity; and the hard wall cylindrical
potential
\begin{equation}
\label{HW}
V_{HW}(\mathbf r_i)=\frac{m^*\omega^2}{2}\frac{[(\delta x_i)^2+
(\frac{y_i}{\delta})^2]^{N}}{r_0^{2(N-1)}},
\end{equation}
where $N$ is a large integer, $r_0$ is the dot radius and $\omega$ is used to
tune the dot to match the harmonic oscillator at $r_0$. The hard wall
potential is not a true step function at $\abs{\mathbf{r}_i}=r_0$, but has a
softness that is decreased with a high $N$ value.

The phonon induced singlet-triplet relaxation rate is calculated from Fermi's
golden rule:

\begin{equation}
\label{GR}
\begin{split}
&\Gamma_i=\\
&\frac{V}{4 \pi^2\hbar} \sum_{j=1}^{3}
\int d^3q \abs{M_j (\mathbf{q})}^2\abs{\bra{S}\hat H_{ph}\ket{T_i}}^2
\delta(\Delta E_{ST} -\hbar c_j q),
\end{split}
\end{equation}
for triplet states $T_{+1}$, $T_{0}$ and $T_{-1}$.
We include relaxation through three phonon effects\cite{PhysRevB.74.035313};
deformation potential coupling
\begin{equation}
\label{M1}
\abs{M_1 (\mathbf {q})}^2=\frac{\hbar \Xi_d^2}{2\rho c_l V}\abs{\mathbf{q}},
\end{equation}
longitudinal piezoelectric coupling
\begin{equation}
\label{M2}
\abs{M_2 (\mathbf {q})}^2=\frac{32\pi^2\hbar e^2 h_{14}^2}
{\epsilon_r^2\rho c_l V}\frac{(3q_xq_yq_z)^2}{\abs{\mathbf{q}^7}},
\end{equation}
and transversal piezoelectric coupling
\begin{equation}
\label{M3}
\begin{split}
&\abs{M_3 (\mathbf {q})}^2=\\
&\frac{32\pi^2\hbar e^2 h_{14}^2}{\epsilon_r^2\rho c_l V}
\left( \frac{q_x^2q_y^2+q_y^2q_z^2+q_z^2q_x^2}{\abs{\mathbf{q}^5}}-
\frac{(3q_xq_yq_z)^2}{\abs{\mathbf{q}^7}}\right).
\end{split}
\end{equation}
The last term is counted twice to account for two identical phonon
modes\cite{PhysRev.136.A869}. Here $\Delta E_{ST}$ is the singlet-triplet energy
splitting, $\mathbf{q}=(q_x,q_y,q_z)$ is the phonon momentum and the states
couple through the $\hat H_{ph}=\sum_{i=1,2} e^{-i\mathbf{q}\cdot\mathbf{r}}$
operator. Linear dispersion is used so that the phonon momentum is matched to
the energy splitting, $\abs{\mathbf{q}}=q=\Delta E /(\hbar c_j)$, where $c_j$ is
the longitudinal speed of sound for  $j=1,2$, and the transversal speed of sound
for $j=3$. $V$ is the normalization volume that will be cancel out, $\Xi_d$ is
the deformation potential constant, $\rho$ is the material mass density and
$h_{14}$ is the piezoelectric constant.

\subsection{Implementation}
The one-particle Schr\"odinger equation, Eq.~(\ref{1eHam}), is solved through
diagonalization of the Hamiltonian matrix
within a  numerical B-spline~\cite{deboor}
basis, where
the polynomial basis allows for  integration to  machine precision through
Gaussian quadrature.
We note that the spin-orbit interaction is included already at this level and
that spin-mixing thus is allowed within the full basis. It has earlier been
shown in the literature\cite{PhysRevB.69.115318,PhysRevB.76.235313}
that a large number of basis functions are needed in order to achieve
convergence for the spin-orbit coupling  even when only  the lowest few
many-body states are of interest.

Using a restricted set of the one-electron states obtained, all possible
two-electron Slater determinants are constructed, and the full configuration
interaction generalized eigenvalue equation is set up and solved
through diagonalization of the two-particle Hamiltonian, matrix,
c.f.~Eq.~\ref{2eHam}. The selection of one-electron orbitals to include is made
by choosing a maximum number of their
$\expect{n_x+n_y}$ quantity and monitoring the convergence in two-electron
energy to within one percent. Here $n_{x/y}$ denote the quantum numbers for the
one dimensional well or harmonic oscillator potential. The $\expect{n_x+n_y}$
quantity is quite constant for a specific state even in the presence of an
elliptic potential and tilted magnetic field. In close proximity to any avoided
crossings created e.g. by the spin-orbit interactions, $\expect{n_x+n_y}$ will
change between the avoiding states.

We model a GaAs dot using bulk values for the effective mass, $m^*=0.067m_e$,
relative permittivity, $\epsilon_r=14.4$ and effective gyromagnetic ratio
$g^*=-0.44$.
For the Dresselhaus parameter, we choose values of;
$\gamma=27.5\text{ eV\AA}^3$, in accordance to previous experimental and
theoretical results in GaAs, \cite{PhysRevLett.90.076807,PhysRevB.59.15882,
PhysRevB.53.3912}; and $\gamma=9\text{ eV\AA}^3$, from previous system specific
results, \cite{PhysRevLett.98.126601,PhysRevB.76.085305,PhysRevB.85.035326}.

For the phonon transition calculations, a crystal density of
$d=5310\text{ kg/m}^3$, deformation potential constant $\Xi_d=6.7\text{ eV}$,
piezoelectric constant $h_{14}=1.4\text{ V/m}$ and sound velocities
$c_L=4720\text{ m/s}$ longitudinal and $c_T=3340\text{ m/s}$ transversal are
employed.

\section{One-Electron Spectrum}
A study of the one-electron spectrum of the potentials will yield some important
information on how the various parts of the one-electron Hamiltonian will
effect the states. The key effect being the spin-orbit interaction and the
avoided crossings appearing when it is included.

\subsection{Harmonic Oscillator Confinement}
\label{sectionharm}
\begin{figure}[ptb]
\includegraphics[width=\columnwidth]{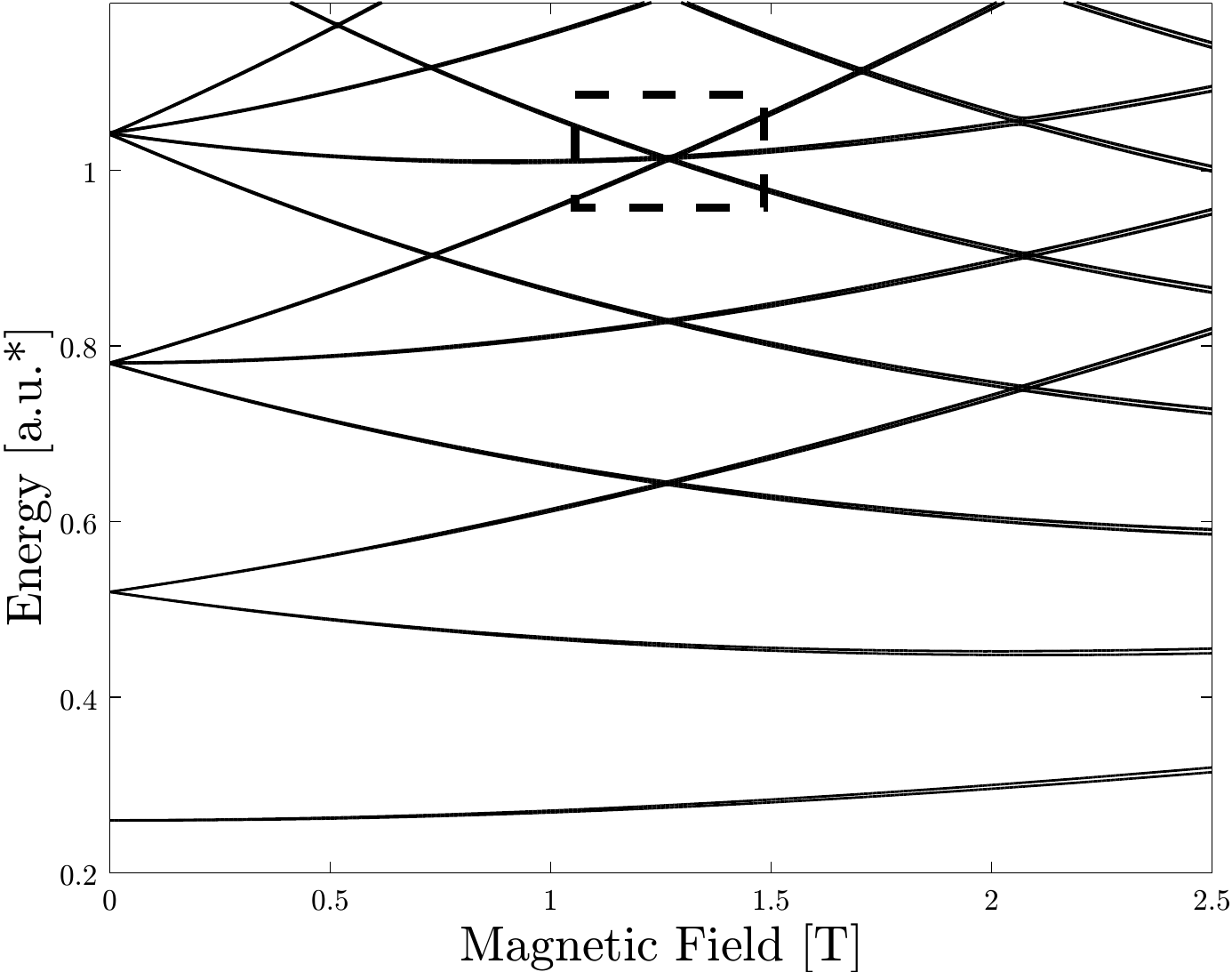}
\includegraphics[width=\columnwidth]{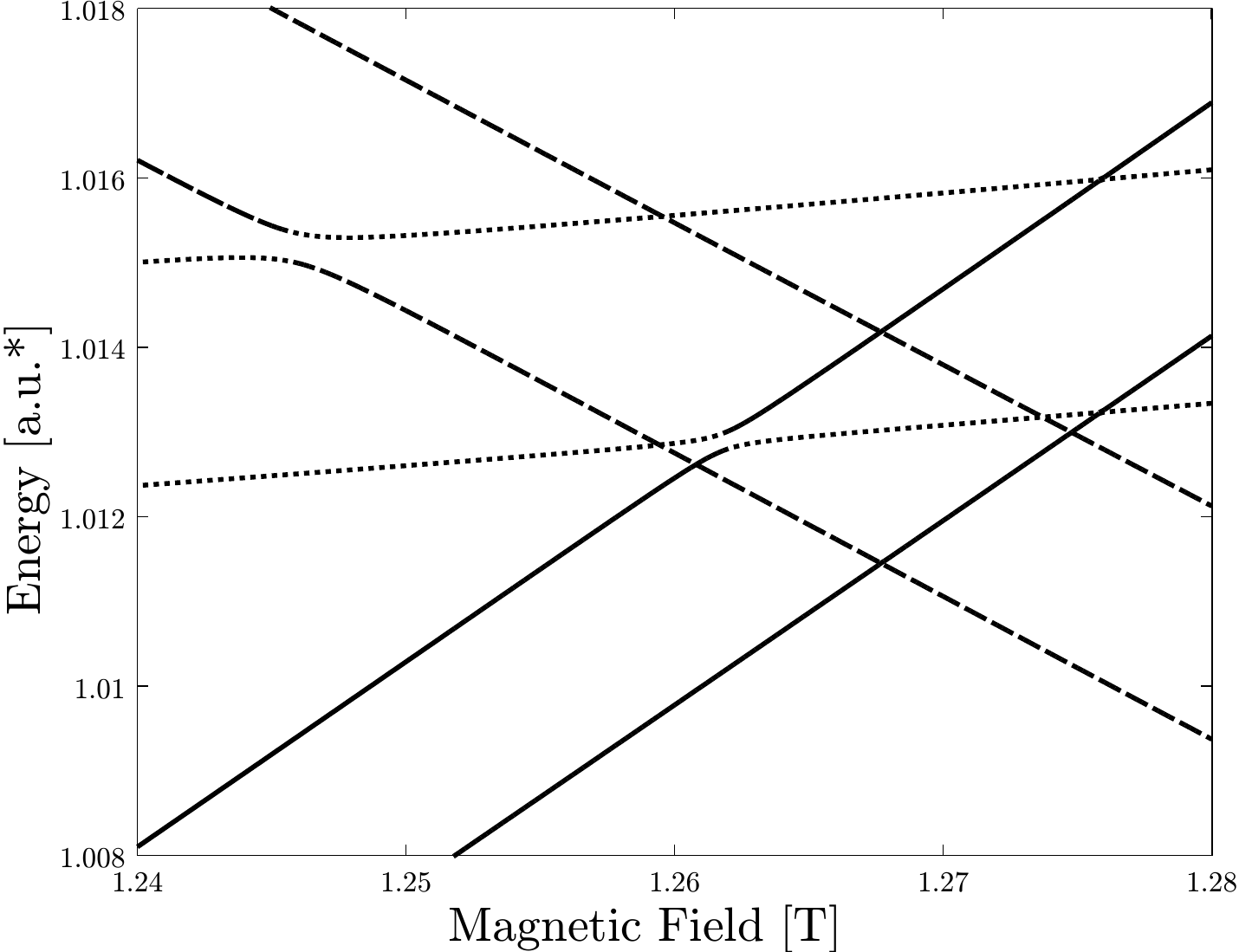}
\caption{Above: The one-electron spectrum of a symmetric harmonic oscillator,
with $\hbar\omega=2.96 \, \text{meV}$, as a function of the strength of a
magnetic field perpendicular to the plane. Below: Enhancement of the dashed area
in the figure above. Avoided crossings due to the cubic Dresselhaus effect seen
between states $\ket{0,+2,\downarrow}\text{(solid)}\leftrightarrow
\ket{1,-1,\uparrow}\text{(dotted)}$ and $\ket{1,-1,\downarrow}\text{(dotted)}
\leftrightarrow\ket{0,-4,\uparrow}\text{(dashed)}$.
}\label{1e_ho}
\end{figure}

\begin{figure}[ptb]
\includegraphics[width=\columnwidth]{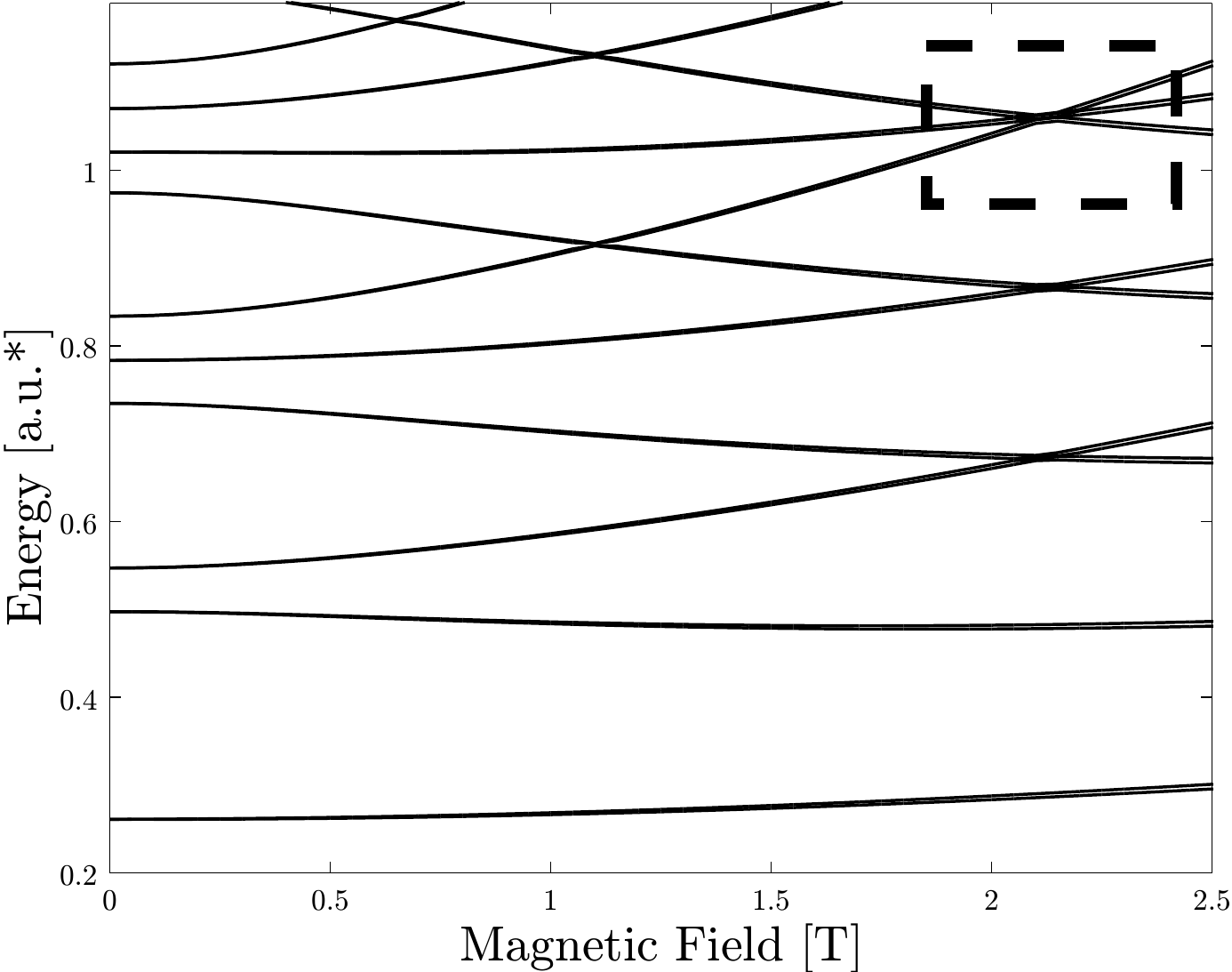}
\includegraphics[width=\columnwidth]{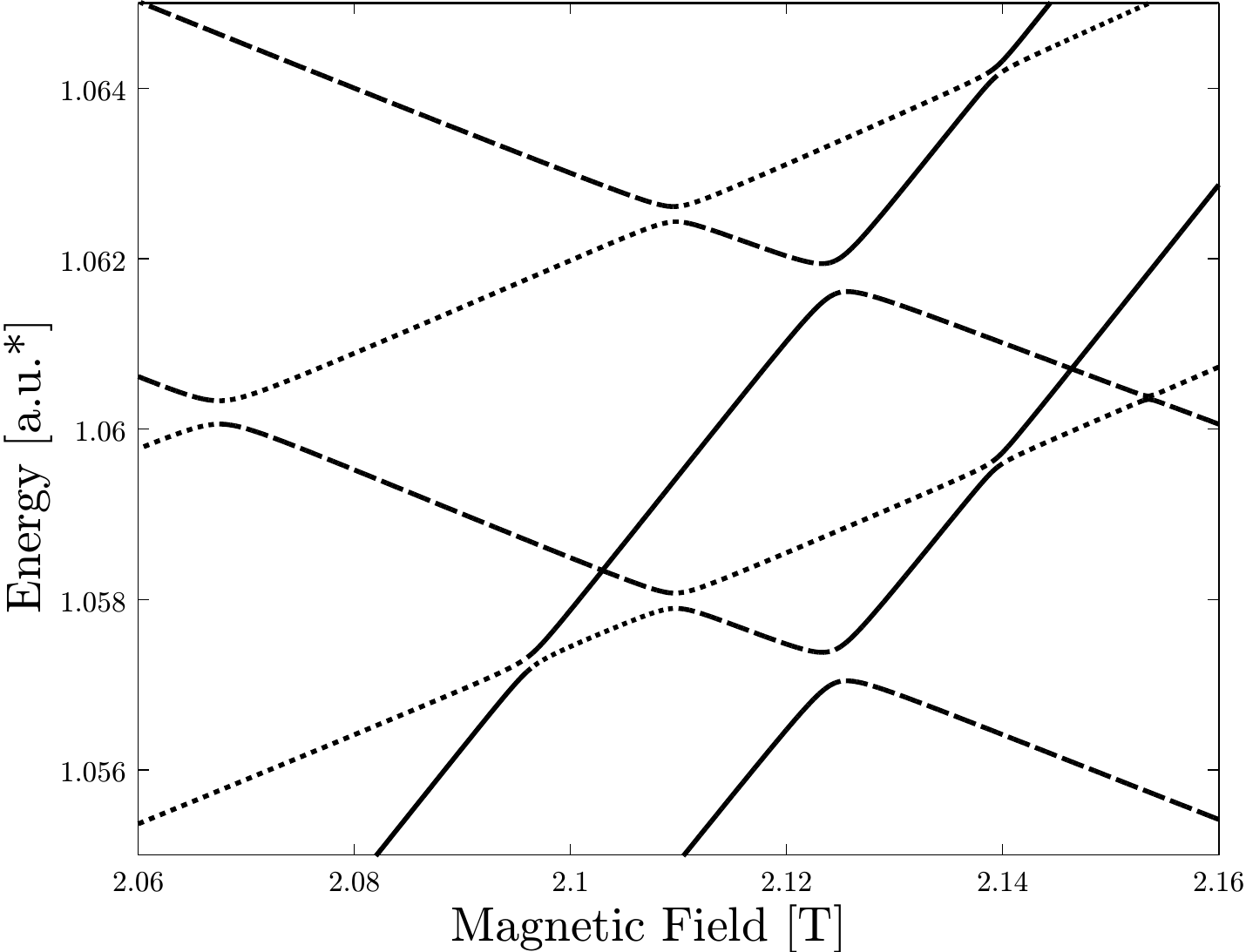}
\caption{Above: The one-electron spectrum of an elliptic harmonic oscillator,
with $\hbar\omega=2.96 \, \text{meV}$ and $\delta=1.1$, as a function of the
strength of a tilted magnetic field, with angles $\theta=55^\circ$ and
$\phi=90^\circ$. Below: Enhancement of the dashed area in the figure
above. Avoided crossings between many states due to the spin states not being
pure $\hat{S_z}$ eigenstates. Line styles as in figure \ref{1e_ho}.}
\label{1e_ho_el}
\end{figure}

The symmetric harmonic oscillator spectrum in Fig. \ref{1e_ho} is familiar
for a magnetic field perpendicular to the plane. The  angular momentum
Zeeman-splitting, linearly dependent on the B-field, dominates at low energies,
but is overshadowed by the $A^2$ term at stronger fields, and at very high field
strengths, Landau levels start forming. An important property of the harmonic
oscillator is the equal-distance level spacing at zero magnetic field. A
consequence of which is that the linear Zeeman splitting will create points
where states bunch together and cross. One such point in Fig. \ref{1e_ho} is
around 1.26 T where states differing by multiples of $\Delta m_l=3$ will cross.

By zooming in on the vicinity of the $m_l=+2\text{, }-1\text{, and }-4$
crossing, we see that the cubic Dresselhaus effect
is creating an avoided crossing
between the $\ket{n,m_l,m_s}=\ket{0,+2,\downarrow}$ and $\ket{1,-1,\uparrow}$
states as well as between the $\ket{1,-1,\downarrow}$ and $\ket{0,-4,\uparrow}$
states, as expected from the selection rules. In the vicinity of these
crossings, the electron spins will be heavily mixed. Since all such
$\Delta m_l=3$ crossings are located at the same magnetic field strength, this
should result in a "spin hot-spot" and
 many-electron states formed from these one-electron orbitals cannot be
eigenstates to $\hat{S}^2$ any more.

The case of the elliptic harmonic oscillator with a tilted magnetic field,
Fig. \ref{1e_ho_el}, has at a first glance a similar spectrum as the symmetric
case. Noticeable differences being an overall scaling since the magnetic
z-component is dampened by a factor $\cos\theta$; and the splitting of states at
low field strength due to the ellipticity.
Since the angular momentum operator, $\hat{L_z}$, is independent of $r$, it
commutes with the circular symmetric harmonic oscillator potential, and it is
possible to choose eigenfunctions that are simultaneously eigenfunctions to the
Hamiltonian and to $\hat{L_z}$. An elliptic potential, however, does not commute
with $\hat{L_z}$. No common eigenfunctions can then be found and
the elliptical states at low magnetic field
strengths are highly mixed in $m_l$, and as a consequence they respond weakly to
the Zeeman effect. Once the magnetic field strength is strong enough to dominate
over the potential, states can once again be approximately described by their
$m_l$ quantum number, and will start to split linearly with $\cos\theta B_0$.

The crossing points seen in the symmetrical harmonic oscillator will however
persist, but at scaled field strengths, since these are a result of the
linear Zeeman splitting that still dominates in this region. The crossing at
1.26 T in Fig. \ref{1e_ho}, has for instance been shifted to 2.11T in Fig.
\ref{1e_ho_el}.

A more detailed investigation of the state
crossings reveals that many more avoided crossings are created by the spin-orbit
interaction in the elliptic case. Since the spin states are a linear combination
of the $\hat{S_z}$ eigenstates, they both will couple to states of
$\Delta m_l=3$. Also second order effects appear, coupling states of
$\Delta m_l=6$. This will result in heavy spin mixing occurring over a broader
magnetic field range, resulting in a larger spin hot-spot than in the  case with
a circular symmetric confinement potential.

\begin{figure}[ptb]
\includegraphics[width=\columnwidth]{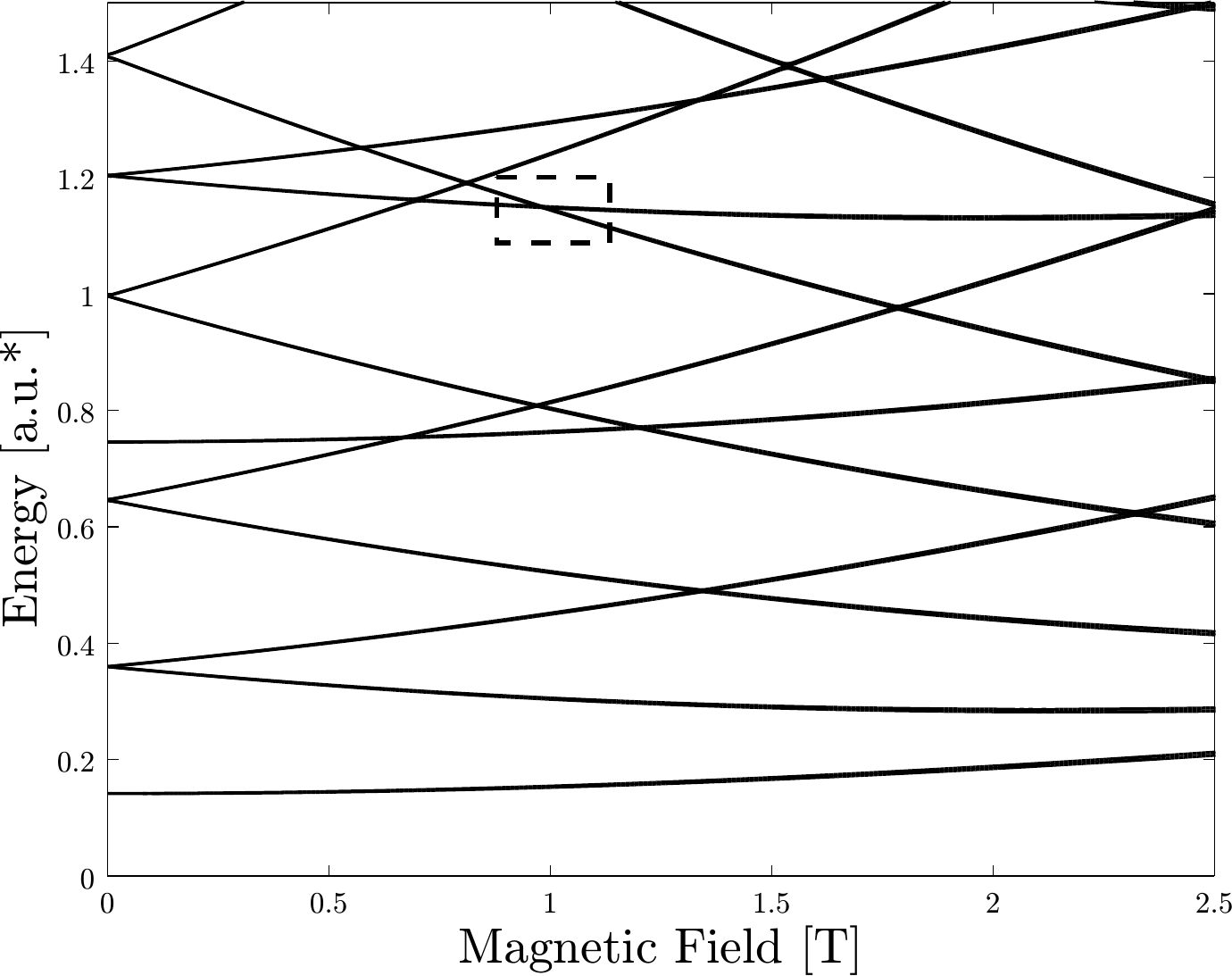}
\includegraphics[width=\columnwidth]{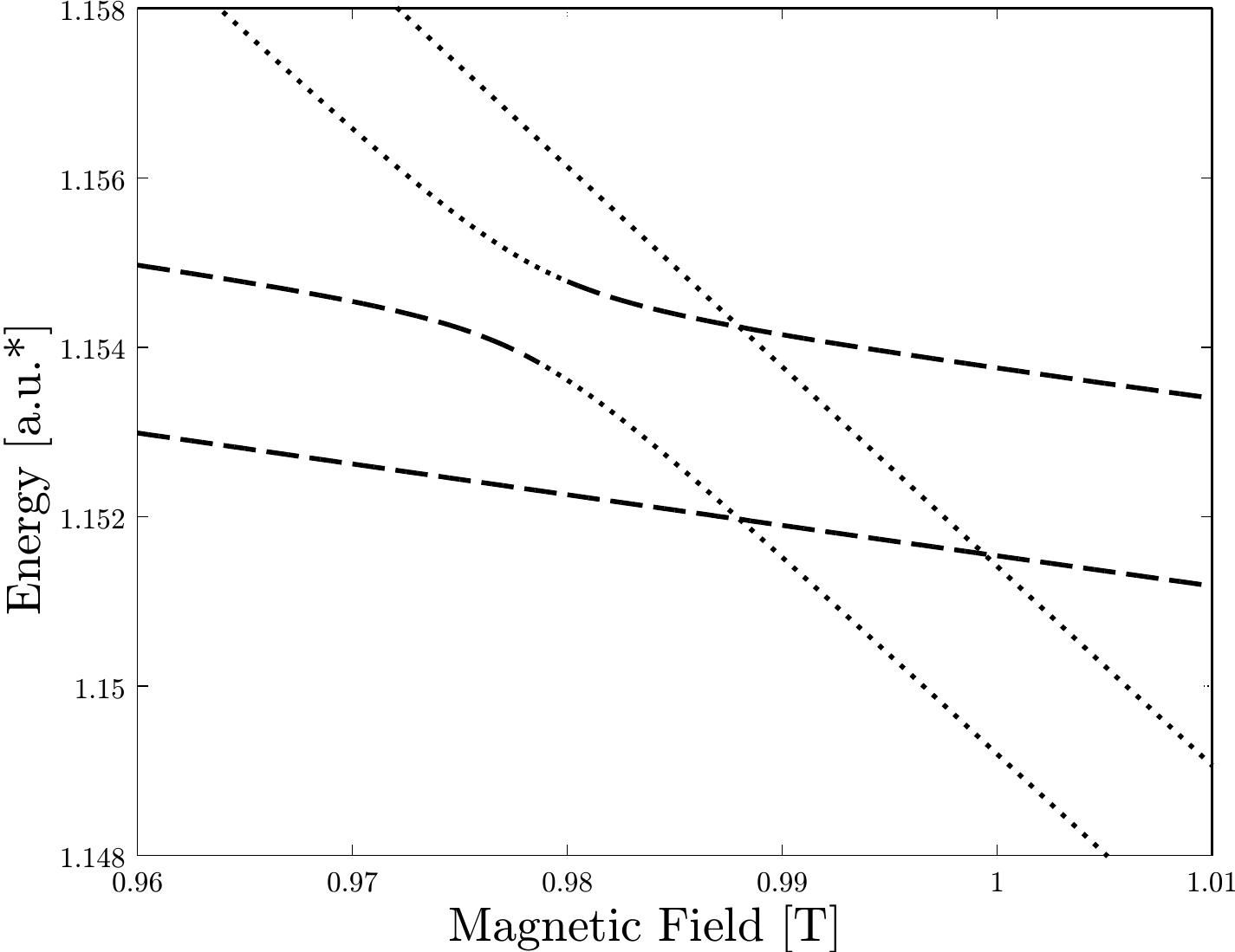}
\caption{Above: The one-electron spectrum of a circular well, with
$r_0=44\text{ nm}$, as a function of the strength of a magnetic field
perpendicular to the plane. Below: Enhancement of the dashed area in the figure
above. Avoided crossings due to the cubic Dresselhaus effect seen between the
$\ket{1,-1,\downarrow}\text{(dotted)}$ and
$\ket{0,-4,\uparrow}\text{(dashed)}$ states.}\label{1e_hw}
\end{figure}

\begin{figure}[ptb]
\includegraphics[width=\columnwidth]{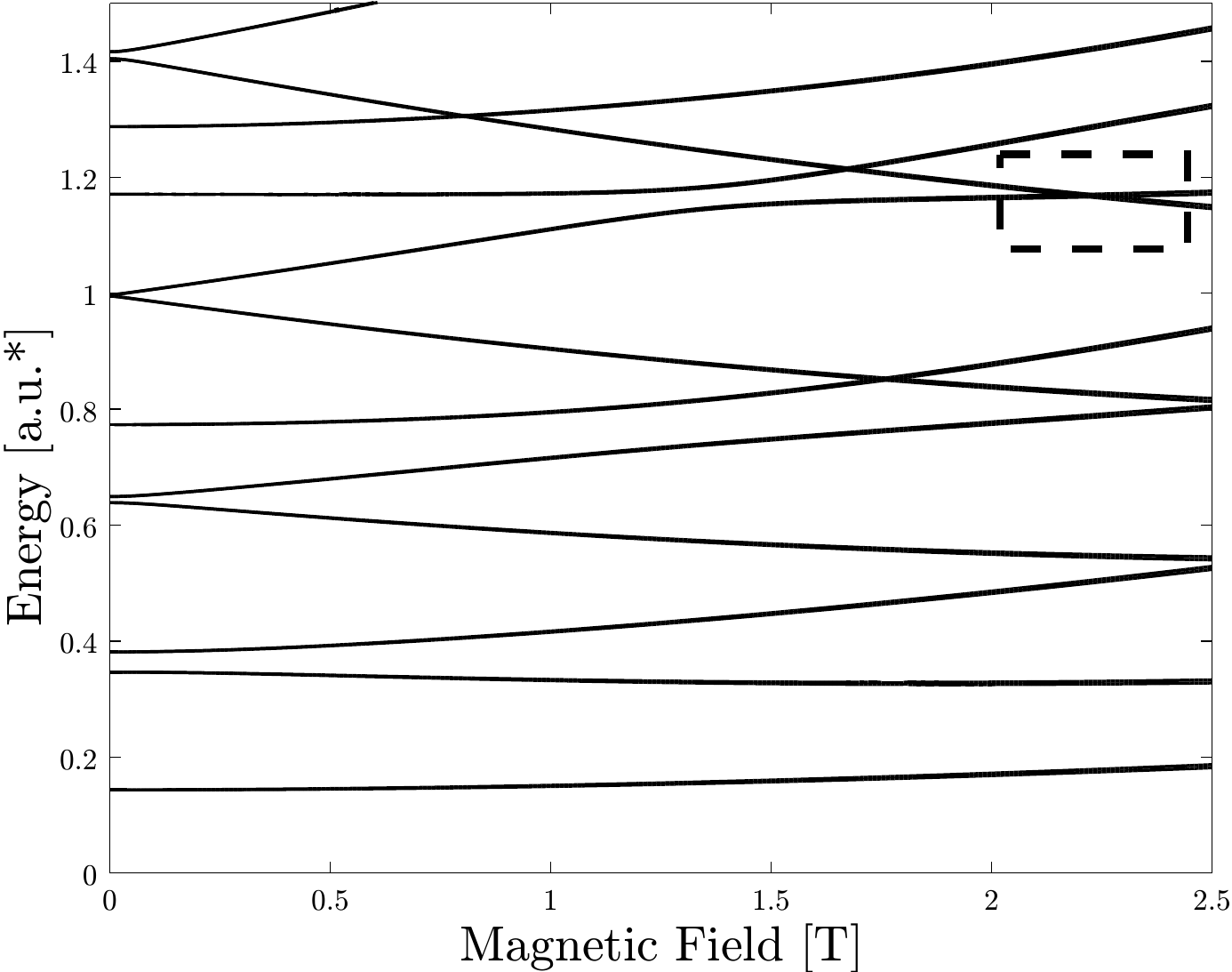}
\includegraphics[width=\columnwidth]{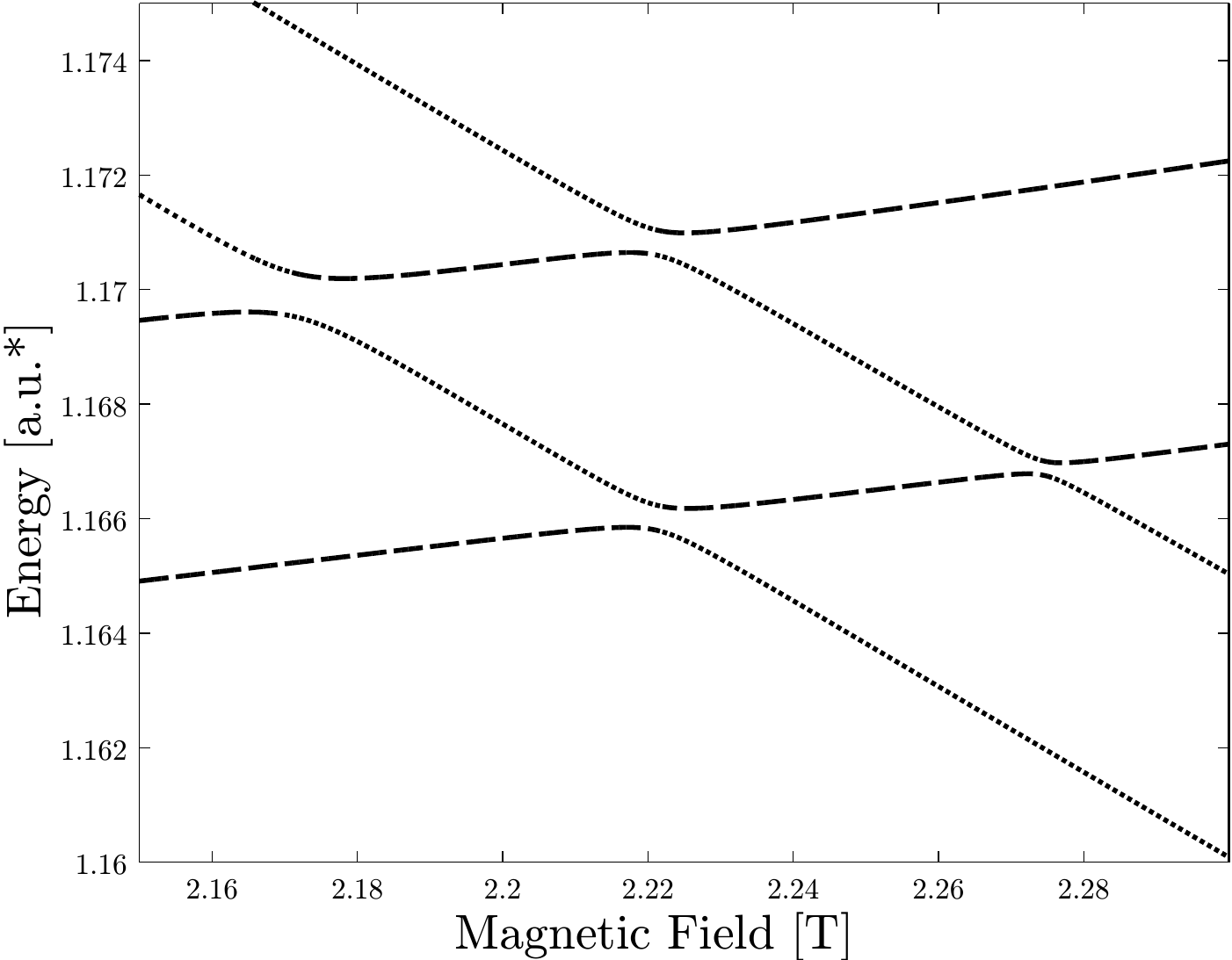}
\caption{Above: The one-electron spectrum of a elliptic well, with
$r_0=44\text{ nm}$ and $\delta=1.05$, as a function of the strength of a titled
magnetic field with angles $\theta=60^\circ$ and $\phi=90^\circ$.
Below: Enhancement of the dashed area in the figure
above. Avoided crossings between several states due to the spin states not being
pure $\hat{S_z}$ eigenstates. Linestyles as in Fig. \ref{1e_hw}}\label{1e_hw_el}
\end{figure}

\subsection{Cylinder Well Confinement}
Fig. \ref{1e_hw} shows the one-electron energy spectra for a circular well
with a magnetic field perpendicular to the plane.
The one-electron state energies at zero magnetic field will differ between the
circular well and harmonic oscillator. Since the electrons in the well are
strictly confined within the radius of the dot, states with higher radial
quantum numbers will have higher energies than their harmonic oscillator
counterparts. The resulting energy level structure, and degeneracies will
therefore differ.

The angular components of the circular quantum well eigenstates are equivalent
to the harmonic oscillator polar angular components, and are hence
eigenfunctions to the angular momentum operator. This results in a Zeeman
splitting for the circular well states, much in the same manner as for the
harmonic oscillator, until the $A^2$-term once again starts to dominate, and
Landau levels begin forming.

An important consequence of not having equidistant energy levels at zero
magnetic field is the lack of high degeneracy at certain field strengths, as
those found in the harmonic oscillator. This should reduce the effect of spin
hot-spots, and instead spread out the spin mixed states over the entire spectra.
In the zoomed in part of Fig. \ref{1e_hw}, an avoided crossing due to the
cubic Dresselhaus interaction can be seen between the $\ket{1,-1,\downarrow}$
and $\ket{0,-4,\uparrow}$ states.

The ellipticity and tilted field added to Fig. \ref{1e_hw_el} produces similar
results to those found in the elliptic harmonic oscillator. The shift due to the
scaled z-component of the magnetic field will not behave quite as linearly
though, and some state crossings will disappear or be shifted far from the
magnetic area of interest.

As in the harmonic oscillator case, the spin states, being linear combinations
of the $\hat{S_z}$ eigenstates, will form avoided crossings with more
states for a tilted magnetic field, as compared to the case with a field
perpendicular to the $xy$-plane.

\section{Two-electron results}
The four lowest two-particle states obtained
by  diagonalization of the two-particle Hamiltonian matrix,
c.f.~Eq.~\ref{2eHam}, are: a singlet state, which is the ground state for modest
B-fields, and the Zeeman-split triplet states, with $M_S =-1, 0, 1$.
The position of the three latter with respect to the ground states is shown as a
function of magnetic field in Fig.~\ref{splitting}.

In the following we want to compare our calculations to the experiment by by
Meunier et al.\cite{PhysRevLett.94.196802}, and the first question is which of
the states that were really addressed there. They claim to populate all three
triplet states, and that the measurements are done on an average over these. We
doubt that this is the case and that the most likely triplet state to populate
should be the lowest energy spin polarized state.
The experiment starts
with a single electron trapped in the dot, a second electron is then allowed to
tunnel into the dot, creating either a singlet or triplet
state. An electron is then allowed to tunnel out and a change in current is
observed with a quantum point contact. Based on tunneling rates
the two-electron state is said to be determined.
With the first electron in a definite state\cite{PhysRevLett.94.196802}, e.g.
the single particle ground state (in the presence of the magnetic field),
$\ket{g,\uparrow}$, the second electron can populate the $\ket{g,\downarrow}$
orbital or the $\ket{e,\uparrow}$ orbital forming  the singlet ground state or
the spin-polarized triplet state, $T_+=\left\{\ket{g,\uparrow}
\ket{e,\uparrow}\right\}$, respectively, where $T_+$ is labeled  by its
dominating configuration. In this case  the $T_-=\left\{\ket{g,\downarrow}
\ket{e,\downarrow}\right\}$ configuration  cannot be reached at all.
Since the true states are better described as a superposition of configurations
they can of course be entered through other configurations, but this is
expected to be a less efficient path.

The $T_0$ state, finally,
should be dominated by a linear combination of the different possibilities there
are to form a $M_S=0$ state from the two lowest orbitals;
$T_0=\frac{1}{\sqrt{2}}\left\{\ket{g,\uparrow}\ket{e,\downarrow}+
\ket{g,\downarrow}\ket{e,\uparrow}\right\}$.
Starting with one electron in a definite spin-state we can only enter this state
through one of the configurations and the state will consequently  be less
efficiently populated. In addition we should expect to form the corresponding
excited singlet state in the case when it is energetically allowed.
This understanding is further supported by later experiments on similar
systems\cite{PhysRevLett.117.236802}, where only one of the polarized triplets
and an unpolarized state is found to be created in addition to the
lowest energy singlet state.

Due to the uncertainty regarding the  individual population of  the three
triplet states, we have chosen to investigate all three states separately rather
than averaging over them.

\subsection{Singlet-Triplet Splitting}

\begin{figure}[ptb]
\includegraphics[width=\columnwidth]{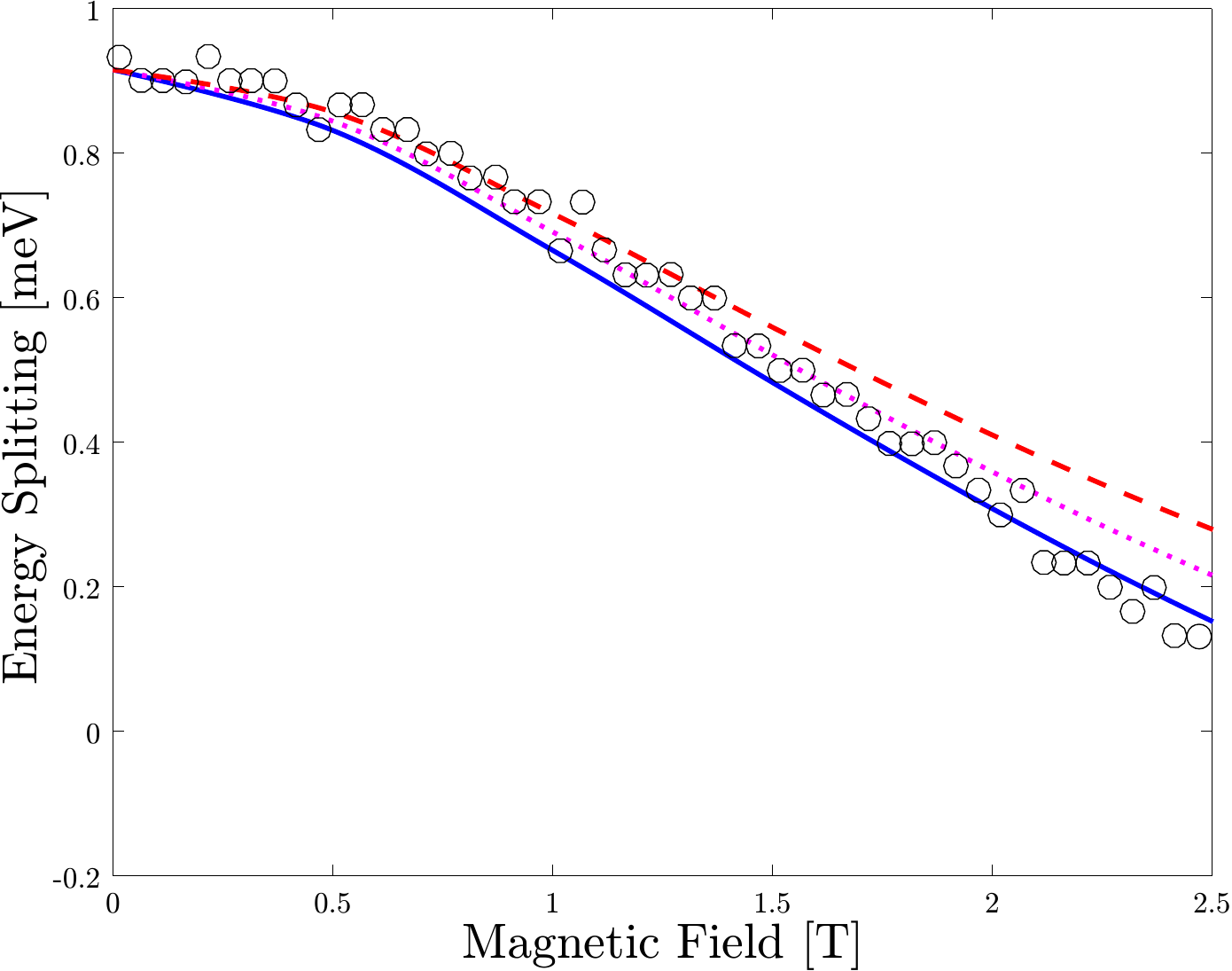}
\includegraphics[width=\columnwidth]{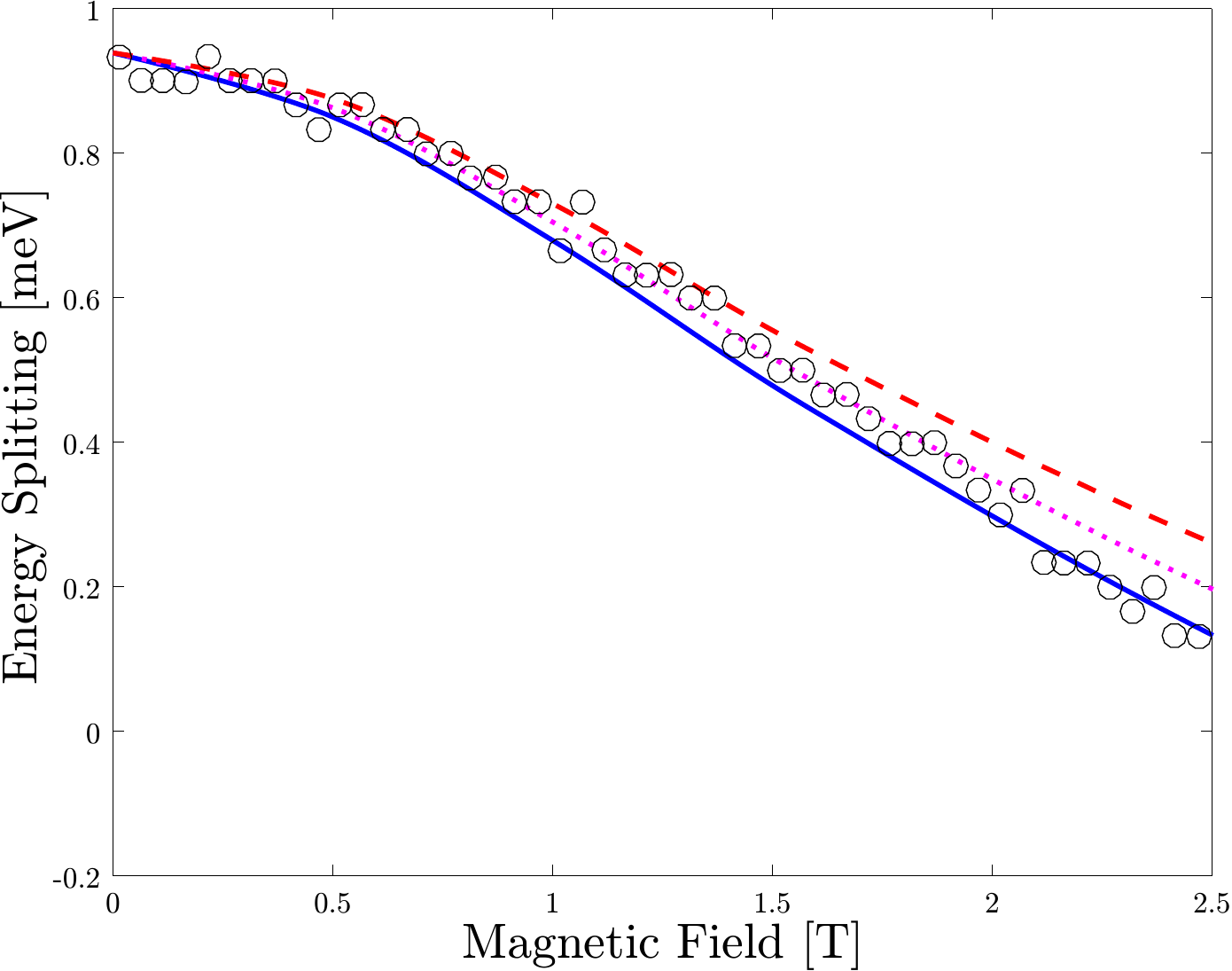}
\caption{
Above: The singlet-triplet splitting of an elliptic harmonic oscillator, with
$\hbar\omega=2.96\text{ meV}$ and $\delta=1.1$, as a function of the strength of
a tilted magnetic field, with angles $\theta=55^\circ$ and $\phi=90^\circ$.
The three triplet states are the first excited $M_S=-1$(solid), $M_S=0$(dotted)
and $M_S=+1$(dashed) states. The circles show the experimental
results\cite{PhysRevLett.98.126601}. Below: The singlet-triplet splitting for an
elliptic well, with $r_0=44$ nm and $\delta=1.05$, as a function of the strength
of a titled magnetic field with angles $\theta=60^\circ$ and $\phi=90^\circ$.
}\label{splitting}
\end{figure}

The experimental results used for comparison\cite{PhysRevLett.98.126601} come
from a system with few parameter details. The inclination of the magnetic field,
measured through Shubnikov–de Haas oscillations, is $68\pm 5^\circ$, with an
unknown azimuthal angle. From the singlet-triplet energy splitting measured and
the constant splitting at low magnetic fields, we estimate the harmonic
oscillator strength to be around $\hbar\omega=3\text{ meV}$ with an ellipticity
$\delta=1.1$, or a well radius of $r_0=44\text{ nm}$ with an ellipticity
$\delta=1.05$. These are rough estimates, but   lead to energy splittings as
those seen in Fig.~\ref{splitting}, which are in good agreement with the
experiment.

If other parameters are used the energy splitting curve will change,
leading to a less good agreement.
For example: First, if the dot is widened, i.e. the radius is increased or the
oscillator strength is decreased,
the Coulomb repulsion between the electrons will decrease. Since
the ground state singlet has a larger correlation energy than the excited
triplet, it will be effected more by the potential change, resulting in a
translation of the energy splitting curve to a lower energy, which will worsen
the comparison with the experimental data.

Second, the ellipticity will both affect the energy splitting at zero field,
and at what field strength the magnetic field starts to decrease the splitting.
As seen in Fig. \ref{1e_ho_el} and \ref{1e_hw_el}, the first excited
one-electron state shifts down in energy when an ellipticity is introduced. This
reduces the total energy of the triplet state since it is dominated by a
configuration where   one of the two electrons occupies this orbital.
The splitting plateau lasts until the one-electron
Zeeman term is large enough to start shifting the energies as discussed in the
previous section. The ellipticity $\delta=1.1$ in the harmonic case, and
$\delta=1.05$ for the well, were found to give the best agreement with the
experimental data.

Third, the dominating magnetic effect in the investigated field range comes from
the linear Zeeman splitting which is directly scaled by the inclination angle
with a factor of $\cos\theta$. Unsurprisingly the inclination angle will then
also affect the singlet-triplet splitting, where a large angle results in a
weaker magnetic dependence. The inclination angle is thus rather strictly
limited to the $55^{\circ}$  and  $60^{\circ}$ found here for the two
confinement potentials.  The in-plane, azimuthal angle will determine the
direction of the elliptic $A^2$ potential, and will shift $m_l$ states somewhat.
It will however be a small effect compared to the effect from the Zeeman
splitting when operating within the field range studied here. We find thus, as
was also concluded previously\cite{Frostad2013}, that the splitting in the
cylindrical well potential matches the experimentally measured inclination angle
better than the harmonic, case although it is still somewhat on the low side.

As a final remark regarding the magnetic field inclination angle,
we note that  both the investigated effective potentials are two dimensional,
there might in addition be some effects from the confinement in the z-dimension
that could better the agreement for both potentials. Inclusion of electron
motion in the z-direction will allow the tilted field to couple not only to the
in plane momentum, but also to the perpendicular one, possibly leading to a
smaller singlet-triplet splitting, but has not been investigated further here.

\subsection{Relaxation}
We now study the relaxation rate as a function of the singlet-triplet splitting
from the previous section.

If only the dominating configurations  are considered for   the triplet states,
one would expect drastically different relaxation times\cite{PhysRevB.75.081303}
for $T_{-1}, T_{0}, T_{1}$, however  the combination of configuration
interaction and spin-orbit mixing has been shown to give much less pronounced
differences for $T_{-1}$ and $T_{1}$\cite{PhysRevB.76.235313}.
Our results are slightly different than what has been found in other studies.
With only the linear Dresselhaus interaction present we find, in agreement with
most of the literature\cite{PhysRevB.77.045328,PhysRevB.73.045304}, that the
$T_0$  state is much more long-lived than the other two, but when also the cubic
Dresselhaus  is included this is no longer the case. We note that
Meunier et al.\cite{PhysRevLett.98.126601}  mention that they do not observe any
slowly  relaxing triplet component in the experiment. More recent
experiments\cite{PhysRevLett.117.236802}, also demonstrate shorter than expected
$T_0$ lifetimes. One possibility  is  that this is due to   a less efficient
population of $T_0$ as discussed above, or that its relaxation is indeed faster
than often believed. Below we discuss the relaxation for the the two different
confinement potentials in more detail.

\subsubsection{Harmonic Oscillator Confinement}

\begin{figure}[ptb]
\includegraphics[width=\columnwidth]{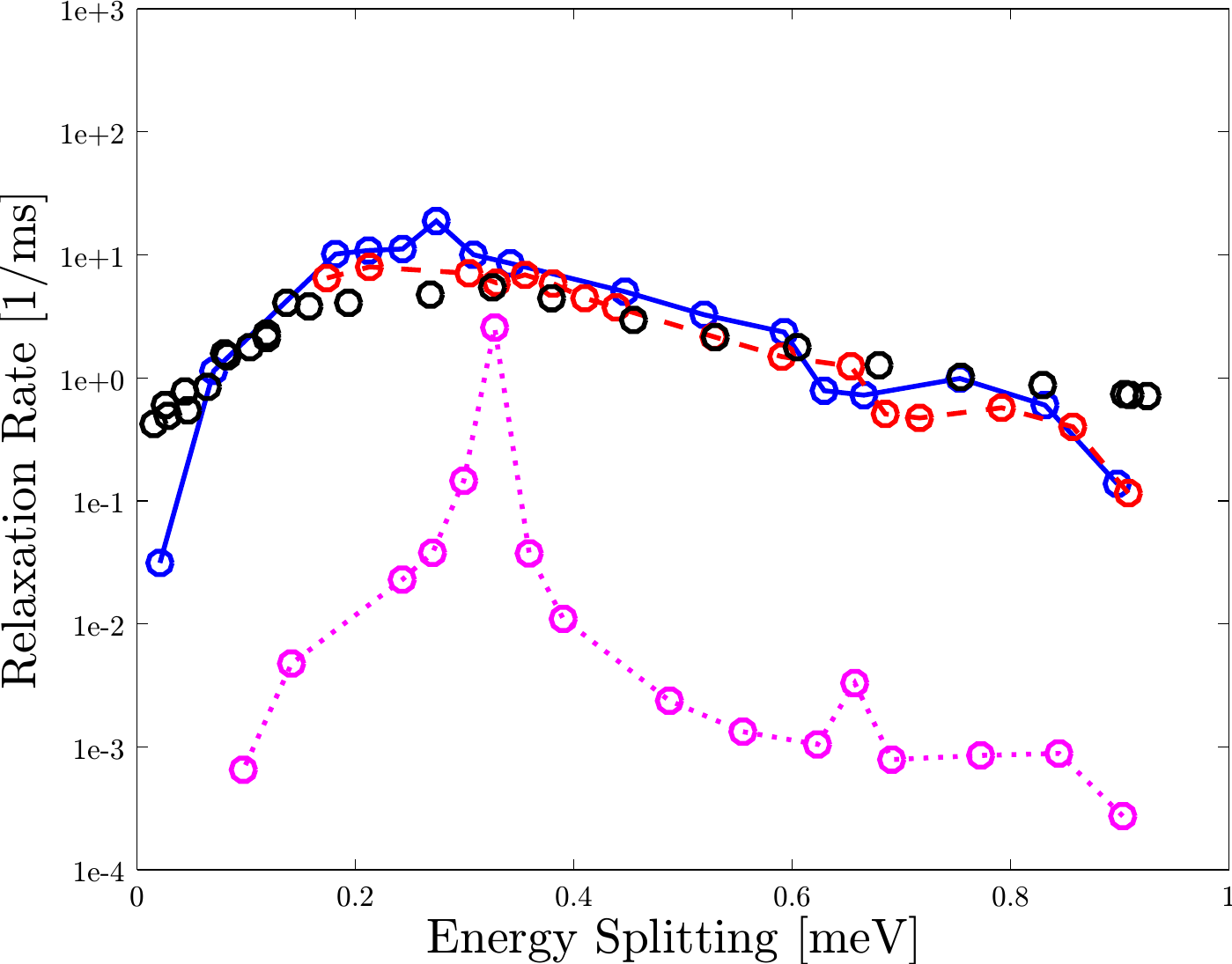}
\includegraphics[width=\columnwidth]{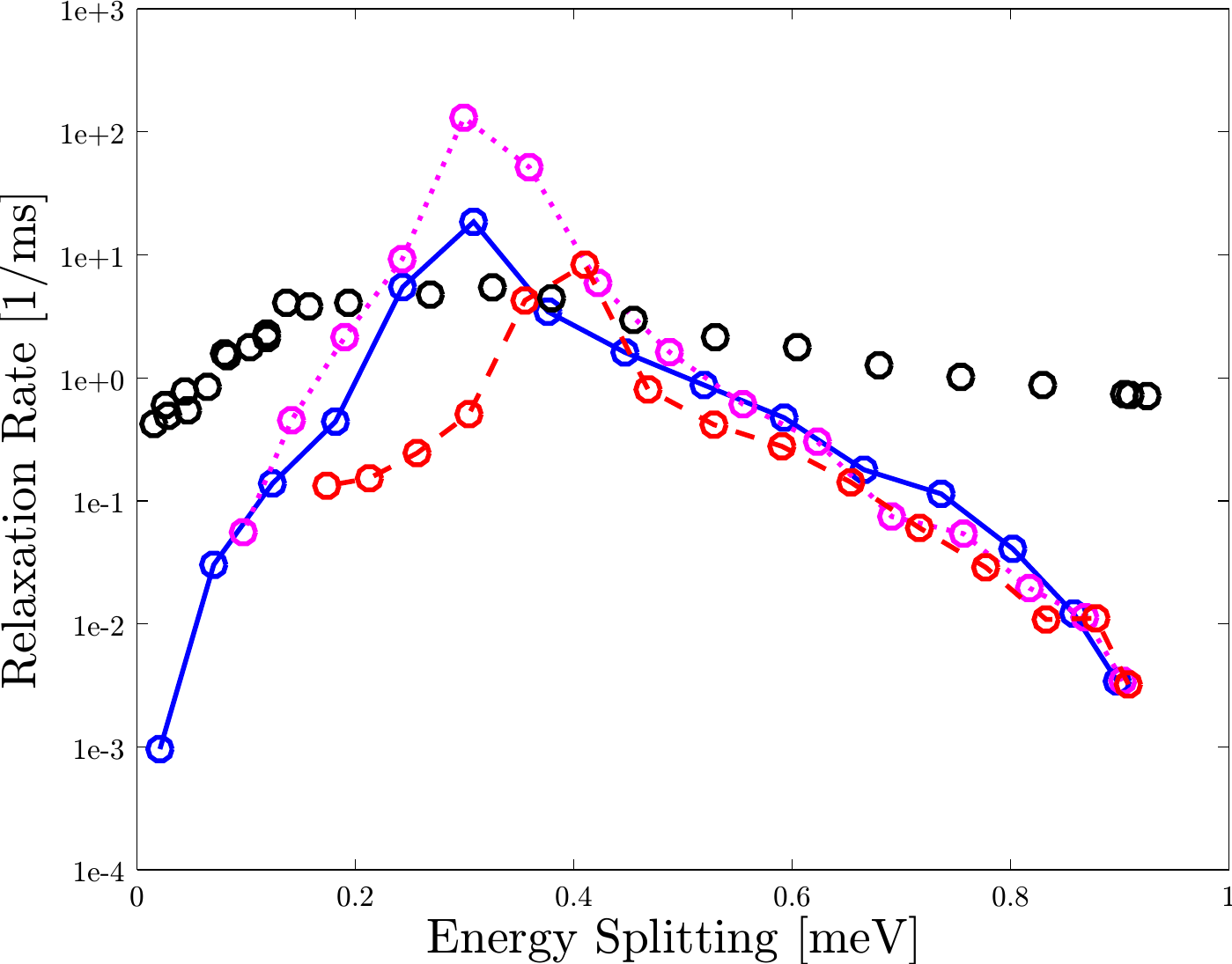}
\caption{
The relaxation rate as a function of the singlet-triplet energy splitting
for an elliptic harmonic oscillator, with $\hbar\omega=2.96\text{ meV}$ and
$\delta=1.1$, in a tilted magnetic field, with angles $\theta=55^\circ$ and
$\phi=90^\circ$. The three triplet states are the first excited $M_S=-1$(solid),
$M_S=0$(dotted) and $M_S=+1$(dashed) states. The circles without
inter-connecting lines show the experimental
results\cite{PhysRevLett.98.126601}. Above: The relaxation rate for
$\gamma=27\text{ ev\AA}^3$, $\hbar\omega_z=11.85\text{ meV}$
 and no cubic Dresselhaus interaction. Below: The rate for
$\gamma=9\text{ ev\AA}^3$, $\hbar\omega_z=11.85\text{ meV}$ with
cubic Dresselhaus interaction.
}\label{ho_relax}
\end{figure}

In the case when only the linear
Dresselhaus interaction is included,
 Fig. \ref{ho_relax}, we find a relaxation maximum around
0.25 meV where the electronic state width matches the wavelength of the phonon.
We are not able to reproduce  the relaxation rate plateau
found at weak fields in the experiment, even
though this type of structure is found in the energy splitting. Possibly some
other relaxation processes dominate at these low field strengths. The relaxation
rates from the spin polarized triplet state, $T_{+1}$ and $T_{-1}$, are fairly
equal in respect to the energy splitting dependence, with the unpolarized $T_0$
state mostly exhibiting a significantly smaller rate, as expected
\cite{PhysRevB.77.045328,PhysRevB.73.045304}. Some structures in the curves seem
dependent on the magnetic field rather than the phonon energy, namely the dip in
rate for the $T_{+1}$ and $T_{-1}$ around 0.65 meV  and the peaks in $T_{+1}$
and $T_{0}$ around 0.3 meV. These points correspond to magnetic field strengths
of 1.1 and 2.1 T, where we also find the crossings of many of the one-electron
basis states  discussed in Sec.~\ref{sectionharm}.

\begin{figure}[ptb]
\includegraphics[width=\columnwidth]{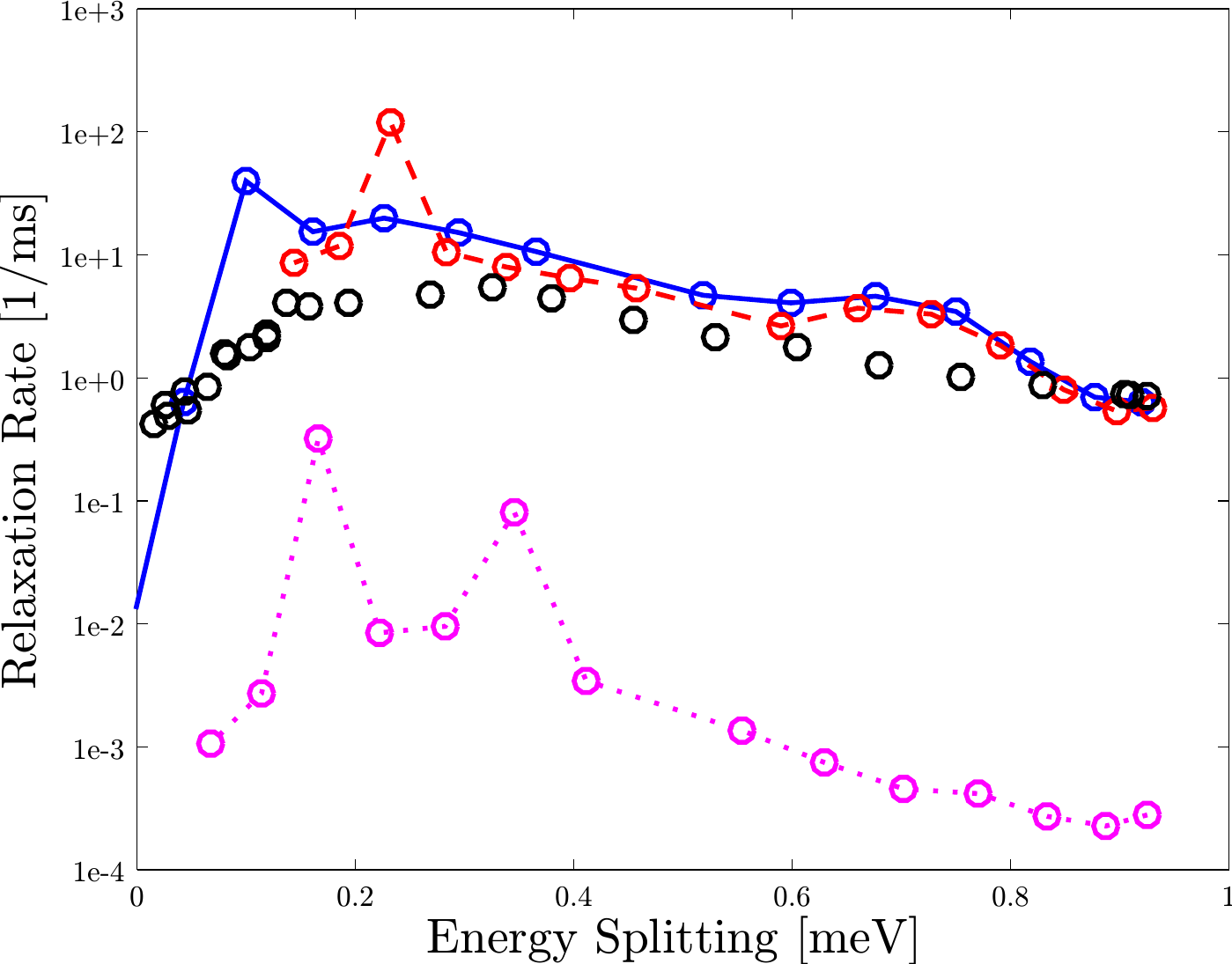}
\includegraphics[width=\columnwidth]{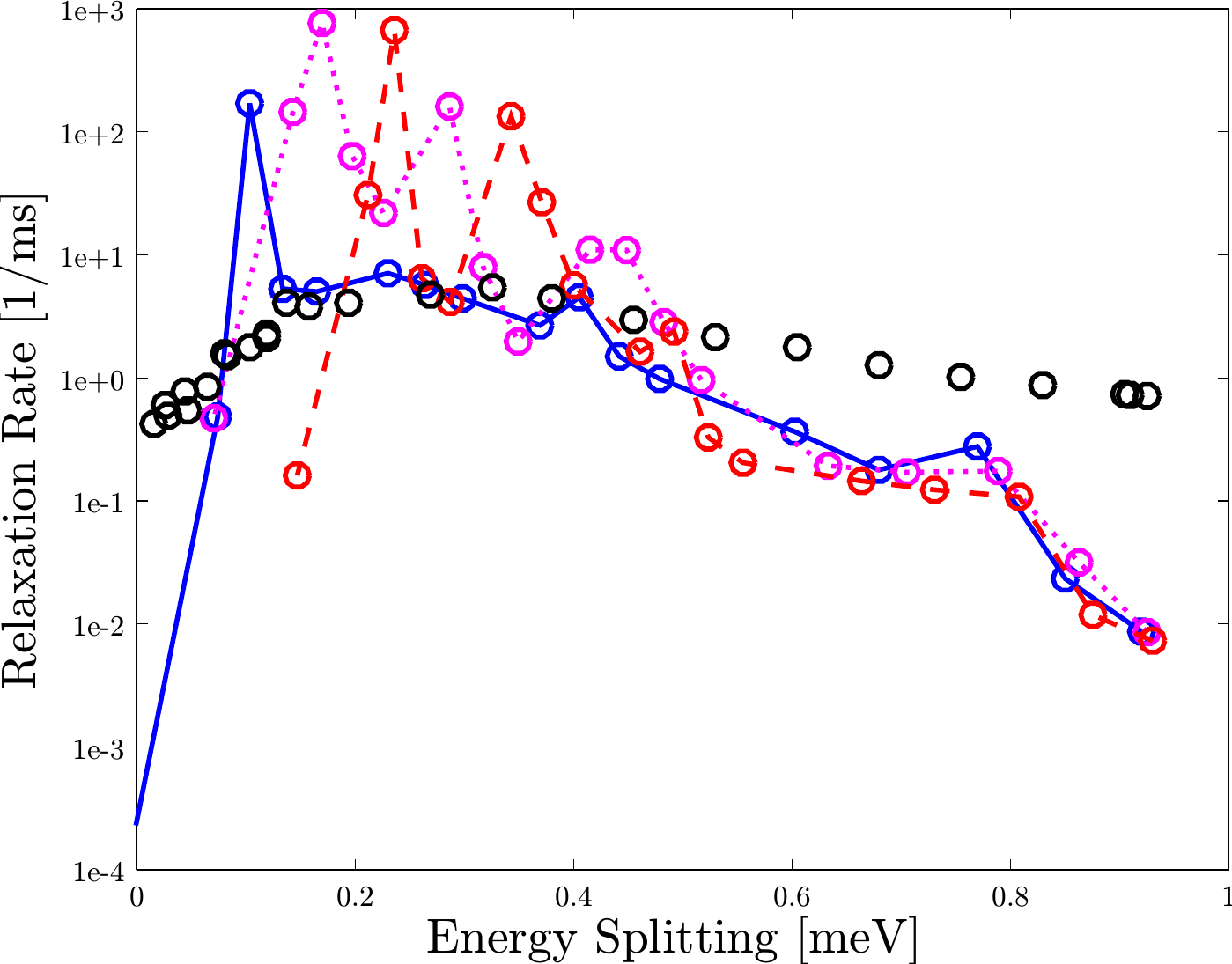}
\caption{
The relaxation rate as a function of the singlet-triplet energy splitting
for an elliptic cylindrical well, with $r_0=44$ nm and $\delta=1.05$, in a
tilted magnetic field, with angles $\theta=60^\circ$ and
$\phi=90^\circ$. The three triplet states are the first excited $M_S=-1$(solid),
$M_S=0$(dotted) and $M_S=+1$(dashed) states. The circles without
inter-connecting lines show the experimental
results\cite{PhysRevLett.98.126601}. Above: The relaxation rate for
$\gamma=27\text{ ev\AA}^3$, $\hbar\omega_z=11.85\text{ meV}$
 and no cubic Dresselhaus interaction. Below: The rate for
$\gamma=9\text{ ev\AA}^3$, $\hbar\omega_z=11.85\text{ meV}$ with
cubic Dresselhaus interaction.
}\label{hw_relax}
\end{figure}

For the calculations where the cubic Dresselhaus has been included, the
parameter $\gamma$ has been tuned down so that the peak in relaxation roughly
matches the experimental data.
The avoided crossings become even more obvious when the cubic
Dresselhaus term is included, where the crossing point of basis states of
$\Delta m_l=3$ becomes a spin hot-spot with a relaxation rate that does not
match anything found in the experimental data. Since the relation between the
linear and cubic Dresselhaus terms is governed by the z-thickness through
$\expect{\pi_z^2}$, a thinner dot should have a increased rate due to a larger
linear contribution\cite{doi:10.1021/jp309389p}. If large enough it should be
able to  overshadow the cubic contribution. A thickness corresponding to an
oscillator strength of $\hbar\omega_z=11.85\text{ meV}$ is used here in
accordance to previous studies
\cite{PhysRevLett.98.126601,PhysRevB.76.085305,PhysRevB.85.035326}. With this
value, the cubic effect seems indeed to be a substantial effect that should be
noticeable in the rates measured. An increased strength of the z-confinement,
which should tune down the relative importance of the cubic effect,
needs however to be compensated by a reduced $\gamma$. In this case it should be
possible to find a better fit to the experimental results, however at a much
lower Dresselhaus coefficient than expected.

In the previous section, to produce the singlet-triplet energy splitting from
the experimental data, a smaller than expected field inclination angle was
required. This shifted the $\Delta m_l=3$ harmonic crossing point from 1.25 T in
the case of a perpendicular magnetic field, to roughly 2.1 T with an inclination
of $\theta=55^\circ$. With the experimentally measured inclination angle of
$\theta=68^\circ$, this crossing should be found
around 3.3 T, beyond the investigated magnetic field range. It is possible that
some three dimensional, or other, effect not included can change the results to
better match the measurements, and through that remove the issue with the cubic
Dresselhaus effect spin hot-spot.

\subsubsection{Cylinder Well Confinement}
The relaxation rate in the case of the cylinder well with only linear
Dresselhaus interaction is similar to the harmonic case, as seen in Fig.
\ref{hw_relax}. The peaks are due to avoided crossings in the one-electron
spectrum, however differently placed than in the harmonic oscillator.
Small structures in the experimental data may be indicative of the underlying
one-electron avoided crossings, these are however not as obvious as in our
calculations. The overall rate is higher than in the harmonic case when only
including the linear interaction, indicating an even smaller Dresselhaus
coefficient.

When the cubic effect is included, the avoided crossing points become more
prominent since the total interaction is increased. More and larger peaks are
present, especially for the $T_{0}$ and $T_{+1}$ states, to the extent that only
the $T_{-1}$ state seems to match the experimental data. The main difference to
the harmonic oscillator is the lack of a large surge in relaxation around the
point of multiple crossings, which does not exist in the cylindrical well.

The difference in relaxation rate between low and high energy splittings is
far larger than in the experimental data, however with a wider and flatter peak
than in the harmonic oscillator case. Altering the modeled thickness of the dot
and adjusting the Dresselhaus coefficient thereafter can possibly improve this.

\section{Conclusion}
A good fit to the experimental singlet-triplet splittings calculated for the two
effective potentials produce similar results for slightly different inclination
angles for the two studied confinement potentials. The hard wall cylindrical
well shows better agreement with the nominal value of  the experiment than the
harmonic oscillator potential, likely due to the stricter electron confinement
in the well potential. The still existing discrepancy between the calculated and
measured inclination angle can possibly be explained by the lacking finite
z-potential in the calculations.

Good agreement for the relaxation rate is achieved when only the linear
Dresselhaus term is included in the computations, although a Dresselhaus
coefficient of  $\gamma=27\text{ ev\AA}^3$ produces somewhat  larger relaxation
rates than the experimental data, but it  can be tuned down slightly for better
agreement. Some magnetic field dependent features can be seen in the
relaxation curves, where the avoided crossings in the one-particle spectrum
occur. These peaks are thin and fairly weak, and may be hard to detect in the
experimental results.

A main finding in the present study is that inclusion of
the cubic term significantly changes the results.  For the harmonic confinement
the results are clearly at odds with the experiment. For hard wall confinement
the spin polarized $T_{-1}$ still shows some resemblance to the experiment
while the other states are too affected by the avoided spin crossings in the
one-electron spectra, creating a large peak in relaxation around these points.
The qualitative differences between the two potential shapes indicate that the
choice of potential may be very important in modeling few electron quantum dots.
Altering the z-confinement may reduce the influence of the cubic term, but will
need to be compensated with a smaller Dresselhaus coefficient.

\begin{acknowledgments}
Financial support by the Swedish Research Council (VR), Grant No. 2016-03789,
is gratefully acknowledged. We would also like to thank Professor Jan Petter
Hansen and Professor Esa R\"as\"anen for providing data and helpful discussions.
\end{acknowledgments}

\end{document}